\documentclass[superscriptaddress,twocolumn,aps,pra]{revtex4-1}
\usepackage{amsmath,amssymb,graphicx,hyperref}

\begin{document}

\title{Far-field head-media optical interaction in heat-assisted magnetic recording}

\author{Ruoxi Yang} \email{Corresponding author: ruoxi.yang@seagate.com}
\affiliation{Seagate Technology, 7801 Computer Avenue S, Bloomington, MN 55435}
\author{Paul Jones}
\affiliation{Seagate Technology, 1000 S Milpitas Boulevard, Milpitas, CA 95035}
\author{Timmothy Klemmer}
\affiliation{Seagate Technology, 1000 S Milpitas Boulevard, Milpitas, CA 95035}
\author{Heidi Olson}
\affiliation{Seagate Technology, 7801 Computer Avenue S, Bloomington, MN 55435}
\author{Deming Zhang}
\affiliation{Seagate Technology, 7801 Computer Avenue S, Bloomington, MN 55435}
\author{Tyler Perry}
\affiliation{Seagate Technology, 7801 Computer Avenue S, Bloomington, MN 55435}
\author{Werner Scholz}
\affiliation{Seagate Technology, 7801 Computer Avenue S, Bloomington, MN 55435}
\author{Huaqing Yin}
\affiliation{Seagate Technology, 7801 Computer Avenue S, Bloomington, MN 55435}
\author{Roger Hipwell}
\affiliation{Seagate Technology, 7801 Computer Avenue S, Bloomington, MN 55435}
\author{Jan-Ulrich Thiele}
\affiliation{Seagate Technology, 1000 S Milpitas Boulevard, Milpitas, CA 95035}
\author{Huan Tang}
\affiliation{Seagate Technology, 1000 S Milpitas Boulevard, Milpitas, CA 95035}
\author{Mike Seigler}
\affiliation{Seagate Technology, 7801 Computer Avenue S, Bloomington, MN 55435}

\begin{abstract}
We have used a plane-wave expansion method to theoretically study the far-field head-media optical interaction in HAMR. For the ASTC media stack specifically, we notice the outstanding sensitivity related to interlayer's optical thickness for media reflection and magnetic layer's light absorption. With 10-nm interlayer thickness change, the recording layer absorption can be changed by more than 25\%. The 2-D results are found to correlate well with full 3-D model and magnetic recording tests on flyable disc with different interlayer thickness. 
\end{abstract}

\maketitle
\date{}

\section{Introduction}
The light delivery subsystem in heat-assisted magnetic recording (HAMR) \cite{Kryder2008Heat} is a key element for the development of high areal-density magnetic recording based on laser-induced local heating. Although a near-field transducer (NFT) \cite{Challener2009Heatassisted,Seigler2008Integrated,Stipe2010} is usually hailed as the ultimate light source beyond the diffraction-limit to achieve deep subwavelength laser spot for heating the magnetic recording layer locally, to efficiently bridge the long distance between laser source and the NFT in micro-scale, an integrated light-path needs to be incorporated to ignite the light-focusing ability of NFTs. With limited coupling efficiency between up-stream light and NFTs, there will usually be background light delivered to the air-bearing-surface (ABS) and transmitted to media. This part of optical power, which is not transferred to the excitation of NFTs through photon-plasmons coupling, is generally far-field in nature compared to the near-field 'radiation' from NFTs. Therefore, the far-field light-media interaction is an important aspect of HAMR optics and has great impact on HAMR design optimization, such as the relation between media-reflection and laser instability, the background light heating of media, etc.

Reflection due to optical waveguide discontinuity has been studied extensively when an additional interface is introduced in the light propagation direction \cite{Ikegami1972Reflectivity,Vassallo1983Orthogonality,Vassallo1988Reflectivity,
Smartt1993Free,Smartt1993Exact,Smartt1994Exact}. In HAMR, the head-media interface has very specific and unique property due to its magnetic recording media design \cite{Rottmayer2006HeatAssisted}. Previous study of waveguide interface is usually discussed within a context of either mirror or anti-reflection coating \cite{Orobtchouk1997Analysis,Nguyen2006Analysis}, optical junction \cite{Smartt1993Exact}, or laser facet \cite{Kendall1993New,Kendall1993Semiconductor}, etc. In short, conventional coatings of waveguides are normally pure dielectrics or pure reflectors. On the contrary, HAMR media is a complex multi-layer system with both dielectric and metal-like materials, which demands additional understanding beyond prior arts. With most available pure-numerical modeling tools taking considerable resource and revealing limited physical insights, a theoretical or semi-analytical approach is of special value for HAMR media study. Due to the existence of extremely thin layers ($\le$ 5 nm) in HAMR media stack, the computation time of finite-element method (FEM) or other finite-difference method can be unnecessarily long and inaccurate. In this paper, we follow a semi-analytical approach combining the plane wave expansion (PWE) and transmission line (TL) theory, and apply this method to understand the far-field optics of HAMR media. We give special attentions to the media thickness sensitivity regarding media reflection and magnetic layer absorption because of their unique significance in HAMR optics. We have also used rigorous numerical methods as well as spin-stand recording tests to verify major conclusions.

The paper will be structured as followed. In Section \ref{sec:theory}, we will briefly review the theoretical aspects of the PWE technique used for light-media interaction study. We then summarize both analytical and numerical modeling results in Section \ref{sec:modeling}, and experimentally verify the modeling predictions in Section \ref{sec:experimental}. Finally we give a recapture of our work followed by acknowledgment. 

\section{Description of Theory}
\label{sec:theory}

\begin{figure}[htbp]
\centering
\fbox{\includegraphics[width=\linewidth]{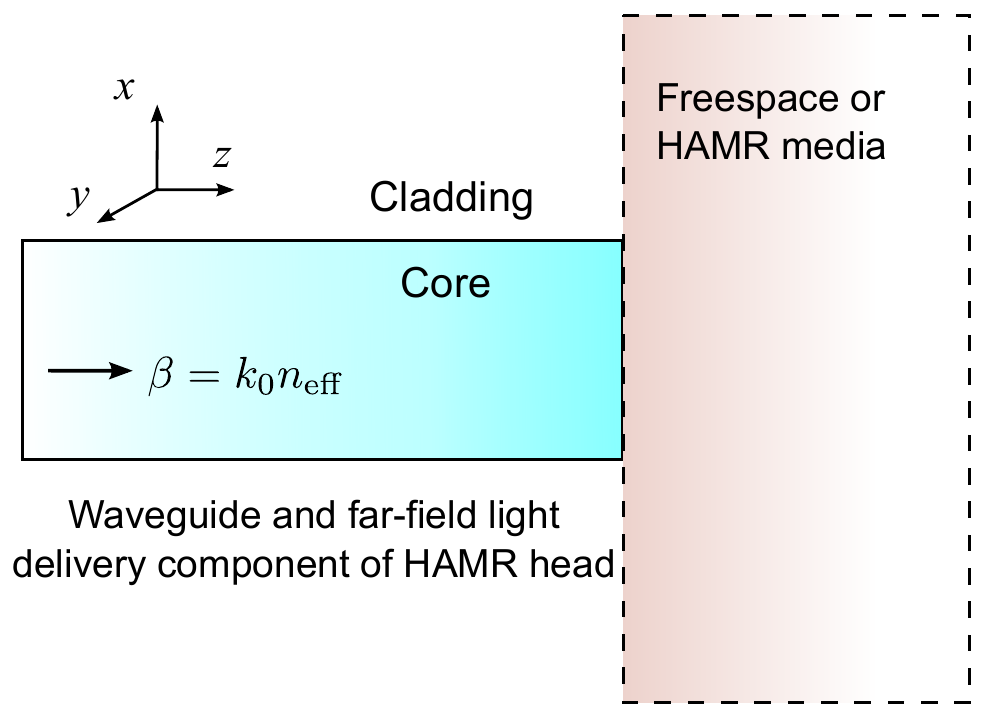}}
\caption{HAMR head-media interface.}
\label{fig:fig1_wvg}
\end{figure}

In this section, we review the technical details of the plane wave expansion (PWE) method. The implementation of PWE in spatial frequency domain follows the general spectral-index principles summarized in \cite{Orobtchouk1997Analysis,Nguyen2006Analysis}. We stick to most of the notations seen in \cite{Kendall1993New,Nguyen2006Analysis}, which have used \textit{s}-space for the spatial frequency Fourier analysis. The coordinate system and the problem definition is sketched in Figure \ref{fig:fig1_wvg}. For 2-D waveguides with guided wave propagating in +$z$ direction, the scalar field in the transverse direction can be expressed as
\begin{equation} 
 \phi_y(x,z)=\phi(x)\exp(-j{\beta}z)
 \label{eq:wave}
\end{equation}
with $\phi$ being either $H$ or $E$ profile for transverse-magnetic ($\mathrm{TM}$) or transverse-electric ($\mathrm{TE}$) respectively and $\beta$ the guiding mode's propagation constant.

Define the Fourier transform of $\phi(x)$ along the $x$ direction in $s$-space as $\hat{\phi}(s)$. The scalar field can therefore be expressed as an integration over all the available $s$ as
  \begin{align}   \label{eq:confft}
   \phi(x)&=\mathrm{FT}^{-1}[\hat{\phi}(s)] \nonumber \\
    & = \int^{+\infty}_{-\infty} \hat{\phi}(s)\exp(-jsx) \,ds 
  \end{align}
where $\hat{\phi}(s)$ gives the complex amplitude of each plane wave with spatial frequency $s$. The equivalent wave number for the plane wave with $\hat{\phi}(s)$ as its amplitude is determined by 
\begin{equation} \label{eq:mom}
	k(s)=\sqrt{s^2+\beta^2}
\end{equation}
and its concept has been discussed in free-space radiation modes (FSRM) methods \cite{Smartt1993Free} and summarized in \cite{Orobtchouk1997Analysis,Nguyen2006Analysis}. It is convenient to obtain the incident angle $\theta$ (relative to the facet normal) as 
\begin{equation} \label{eq:angle}
	\theta_i=\tan^{-1}\frac{s}{\beta}
\end{equation}

For continuous $x$ and $s$ used in analytical expressions, the incident and reflected wave are given by Eq.~\ref{eq:confft} and 
\begin{equation} \label{eq:confftR}
   \phi_r(x)=\int^{+\infty}_{-\infty} r(s)\hat{\phi}(s)\exp(-jsx) \,ds
\end{equation}
in which $r(s)$ is the $s$-dependent transverse field reflectivity. For TM polarization, the $E$-field reflectivity calculated from cascaded TL theory should be converted back to $H$-field reflectivity by adding a minus sign, unlike the case of TE. The discretization of Eqs.~\ref{eq:confft} and~\ref{eq:confftR} for N sampling points can be written as
\begin{equation} \label{eq:disfft}
   \phi_i[x_j]= \sum_{i=1}^{N} {\hat{\phi}[s_i]}\exp{(-js_{i}x_{j})}
\end{equation}
and
\begin{equation} \label{eq:disfftR}
	\phi_i[x_j]= \sum_{i=1}^{N} {r[s_{i}]\hat{\phi}[s_i]}\exp{(-js_{i}x_{j})}
\end{equation}

Here $x_{_j}$ belongs to the discrete $x$-space. The complex reflectivity $r(s)$ is calculated by the generalized TL theory. We adopted the general approach described in \cite[chaps. 8 and 9]{ewa} to implement the TL calculation for lossy multilayer after getting the field profile of the dielectric waveguide via transcendental equations. Starting from the effective index of lightwave with vacuum wavelength $\lambda_0$, with the incident angle and refractive index of the first medium ${n(s)={k(s)}/{k_0}}$ known (${k_0={2{\pi}}/{\lambda_0}}$) we have:
\begin{equation} \label{eq:neff}
   n_{\mathrm{eff}}=n(s)\sin{\theta_i}
\end{equation}
in which the incident angle $\theta_i$ has been given by Eqs.~\ref{eq:mom} and~\ref{eq:angle}.

The first medium's optical property is given by the spectral index that changes with $s$. The generalized Snell's law is used to determine the refractory angle, which again allows complex values \cite{ewa}. The impedance at each surface is calculated from the right (in terms of the positive z direction) to left recursively. Assuming the $n^\mathrm{th}$ medium of the multi-layer has refractive index as $N[n]$ and thickness as $D[n]$, with a characteristic impedance $Z[n]$. The parameters to determine the field distribution in the $n^\mathrm{th}$ medium can be calculated with reflection coefficient $\Gamma$ as:

\begin{equation}  
 \left. 
 \begin{aligned}
        \Gamma_r[n]= &\frac{Z_l[n+1]-Z[n]}{Z_l[n+1]+Z[n]}\\
        \Gamma_l[n]= &\Gamma_r[n] \exp{(-j2\psi[n])}      
 \end{aligned} 
 \right\}  \label{eq:recur}
\end{equation}
with the impedance given by
\begin{equation}   
		Z_l[n]=Z[n] \cdot  
		\frac{Z_l[n+1]+jZ[n]\tan{(\psi[n])}} {Z[n]+jZ_l[n+1]\tan{(\psi[n])}}      \label{eq:imped}
\end{equation}
where $\Gamma_{r,l}$ denotes the reflection coefficient on the right or left boundary (relative to the +$z$ direction), and $Z_l$ the impedance on the left boundary. Here the longitudinal phase term $\psi[n]$ is obtained from 
\begin{equation} \label{eq:psi}
   \psi[n]={k_0}N[n]D[n]\cos({\theta[n]})
\end{equation}

The reflection of the multilayer system equals to the field reflectivity of the first interface. As noted previously, in convention this is the $E$-field reflectivity for both TE and TM cases. 

\section{Calculation and Modeling}
\label{sec:modeling}

In this section, we will present the modeling results of how media reflection and absorption change with HAMR media thickness variation. We will firstly show the results of media reflection with plane-wave input to verify the TL theory, and then combine TL and PWE to study the waveguide-media interaction in 2-D, after comparing with Finite-Difference Time-Domain (FDTD) results for the simplified waveguide-air case. Finally we show the Finite-Element Method (FEM) modeling results of full 3-D HAMR heating process and discuss its correlation with 2-D data. All the semi-analytical results are based on the generic media stack introduced previously by Advanced Storage Technology Consortium (ASTC) \cite{Rausch2011}. The media stack consists of (from top to substrate) a head coating, an air gap, a media coating, a recording layer, an interlayer, a heatsink layer and a glass substrate. The wavelength of interest in the theoretical part is fixed at 800nm, so as to be consistent with the ASTC table. The general analytical approach given in the paper can be readily applied to other spectra. The optical property and thickness is summarized in Table \ref{tab:astc}, with Layer 4 and Layer 6 representing recording layer and heatsink layer respectively.

\begin{table}[htbp]
  \caption{ASTC Media Stack for 800nm light.}
  \begin{center}
    \begin{tabular}{ccccccccc}
    \hline
    LayerID & 1 & 2 & 3 & 4 & 5 & 6 & 7\\
    \hline
    n    & 1.60 & 1.00 & 1.20 & 2.90 & 1.70 & 0.26 & 1.50\\
    k     & 0.00 & 0.00 & 0.00 & 1.50 & 0.00 & 5.28 & 0.00\\
    D(nm) & 2.5 & 2.5   & 2.5  & 10.0 & 15.0 & 80.0 & inf.\\
    \hline
    \end{tabular}
  \end{center}
  \label{tab:astc}
\end{table}

\subsection{Planewave-Media Interaction}

The interaction between plane-wave and media is the most simple case for light-media interaction. Naive as it may look like, the study has proved to be practically meaningful since most media optical probing methods (including ellipsometer) utilize plane-wave-like light beam as the incident source. In other word, the optical response of any media to a plane wave illumination can be regarded as to some extent intrinsic to the media. In addition, the concept behind PWE for waveguide mode depends on the decomposition of waveguide eigenmodes into multiple plane waves at different incident angles. Thus the behavior of plane-wave on HAMR media provides direct understanding and learning of the HAMR media multilayer optical system. 

\begin{figure}[htbp]
\centering
\fbox{\includegraphics[width=\linewidth]{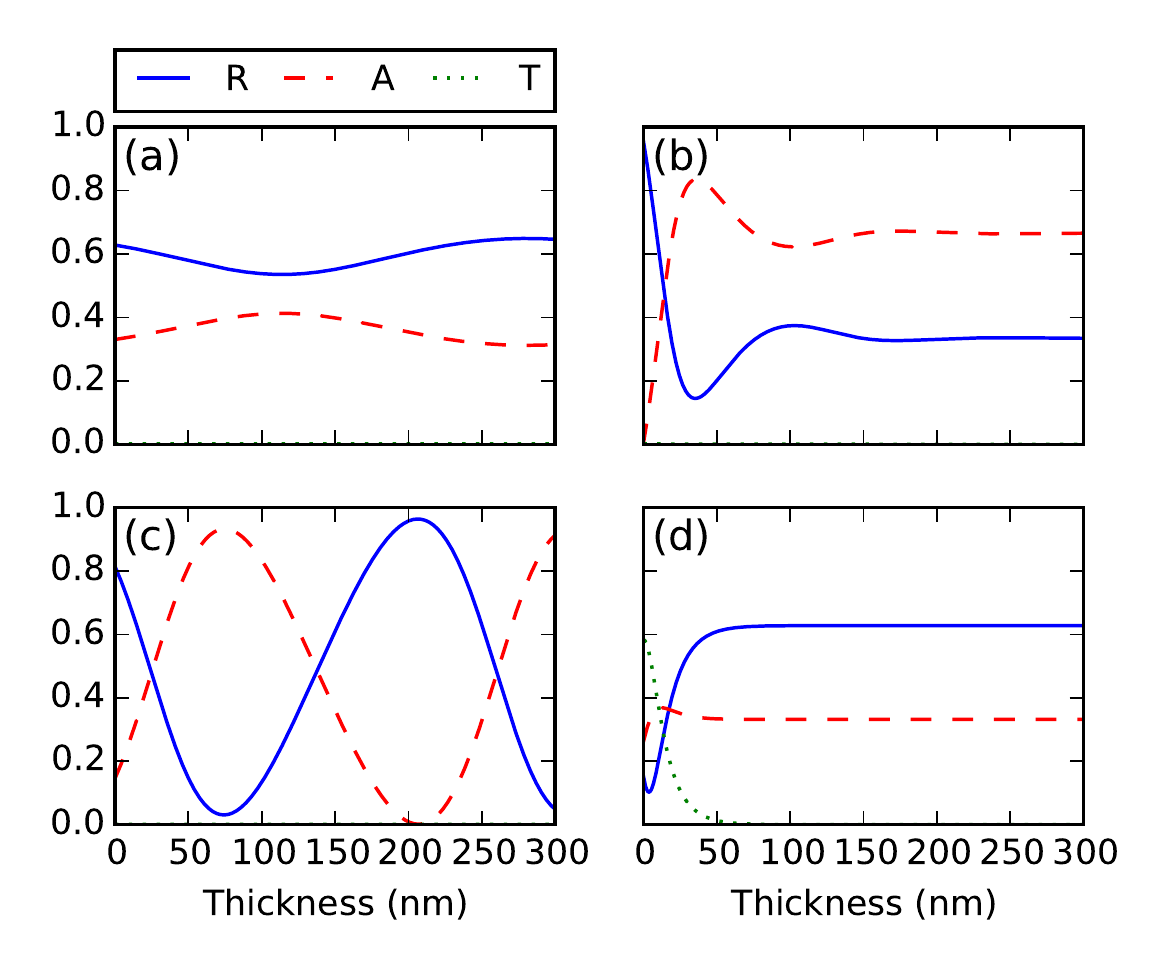}}
\caption{Reflection (labeled 'R', blue solid line), recording layer absorption (labeled 'A', red dashed line) and transmission (labeled 'T', green dotted line) (all normalized to the incident power) vs. different layer thickness, when changing the thickness from 0 to 300nm of (a) overcoat, (b) recording layer, (c) interlayer and (d) heat sink layer.}
\label{fig:swing}
\end{figure}

With the generalized TL theory reviewed in the last section, we fix the incident angle at zero (normal incidence) and sweep the thickness of media overcoat (LayerID=3), magnetic layer (LayerID=4), interlayer (LayerID=5) and heatsink (LayerID=6). The field is launched from the air (refractive index $n$=1.0). In Figure \ref{fig:swing}, we plot the reflection and normalized media absorption vs. thickness curve for the four layers of interest, each from 0 to 300nm. The curve 'R' is the media stack total reflection, 'A' the recording layer absorption and 'T' the stack transmission. Understandably, there is engineering limit for the reasonable range of those layer thickness and except for the heatsink layer, all the media thin-film thickness is of the order of 10nm or even less. Yet from a pure optical point of view, it is highly valuable to appreciate the different sensitivity in media reflection and recording layer absorption. Below are a few salient takeaways from the visualization.

Firstly, we notice that in most cases, increased media reflection is accompanied by the drop of recording layer absorption. Magnetic layer absorption has first-order impact on HAMR heating efficiency and laser power requirement, so the 'A' curve should be considered a crucial gauge for media heating efficiency. One minor exception happens with the heatsink thickness being small, in which the reflection and absorption increases at the same time, yet when the anomaly happens the absorption change is relatively small.  

Secondly, we see that the sensitivity of interlayer thickness is relatively big compared to other layers. Judging from the oscillating and periodic nature seen in Fig. \ref{fig:swing}(c), this effect can be explained almost purely by considering the impedance matching capability of the dielectric layer (interlayer) on top of a good reflector (heatsink). For the ASTC media stack, an absorption peak as well as a reflection valley is seen for interlayer thickness at $\sim$75nm, and the recording layer absorption keeps rising with increased interlayer thickness until reaching the absorption local maximum. As a side note, a damped oscillation can be seen in the recording layer tuning case (Fig. \ref{fig:swing}(b)). The total absorption is almost linear to the recording layer thickness up to $\sim$40nm. Near its nominal thickness ($\sim$10nm), there is no huge improvement in terms of media absorption efficiency (absorption per thickness). 

\begin{figure}[htbp]
\centering
\fbox{\includegraphics[width=\linewidth,page=1]{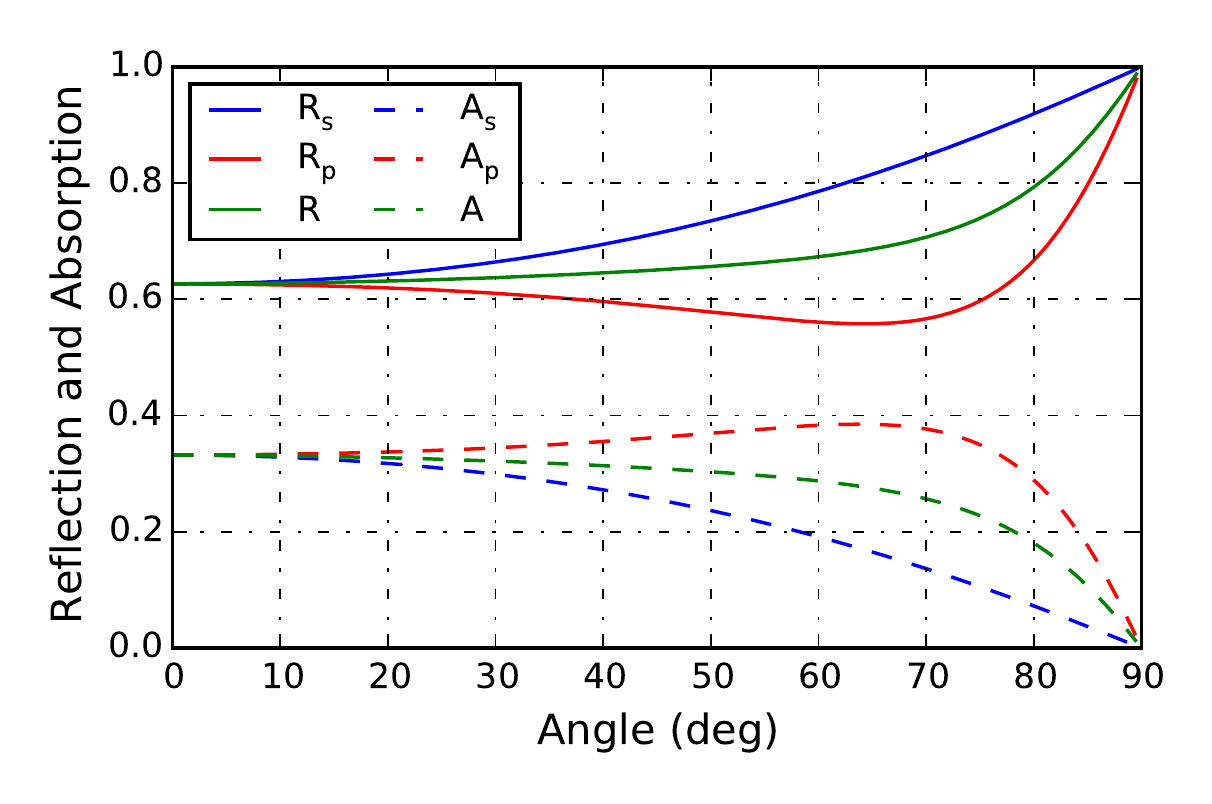}}
\caption{Reflection (R) and recording layer absorption (A) for different polarizations ($s$ or $p$) vs incident angle for ASTC shared media stack. R (or A) without polarization subscript gives the average of $s$- and $p$-wave, which denotes the case of a non-polarized illumination.}
\label{fig:angle}
\end{figure}

Finally, we notice that with heatsink thickness growing, the reflection increases and then saturates, leaving the recording layer absorption unchanged at the upper end. Thus there is relatively limited optical benefits related to further tuning heatsink thickness when it reaches the level of optical skin depth. Also, changing the thickness of a lossless media overcoat barely changes the recording layer absorption. In real head-media interaction, however, it is expected that the head-media spacing will be a highly sensitive factor in determining the heating efficiency since the near-field light becomes the major heat source.

Next we preserve the ASTC media stack's optical properties, while changing the incident angle of the input beam. In Figure \ref{fig:angle}, we plot the angular reflection and absorption for ASTC stack illuminated by a plane-wave. There is a relative reflection valley at about 70 deg for $p$-wave, while no local reflection minimum or absorption peak is seen for $s$-wave. We skip the transmission curve as it is zero with all incident angles due to the 80nm heatsink layer. We have also observed that the $s$-wave has generally more reflection and less magnetic layer absorption especially for large incident angles. This observation suggests a quasi-TM ($p$-like) light delivery system generally couples with HAMR media more easily, although to observe a significant difference between the polarizations the incident angle of the beam needs to be relatively large ($>$40 deg). A more relevant scenario will be presented in the next section, in which the incident beam is launched from high-index waveguide other than from the air. 

\subsection{Waveguide-Media Interaction}

\begin{figure}[htbp]
\centering
\fbox{\includegraphics[width=\linewidth]{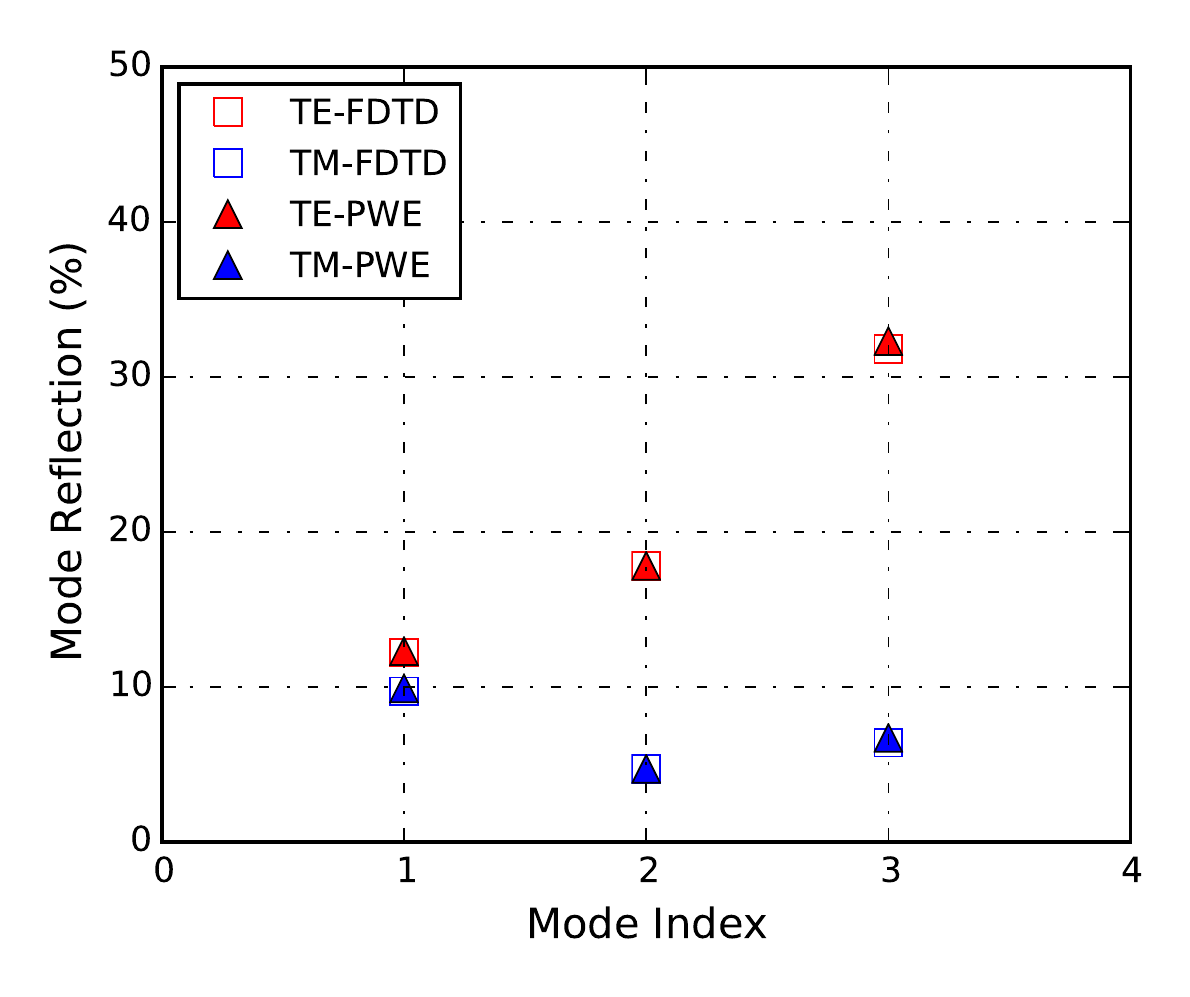}}
\caption{PWE calculated waveguide reflection (mode overlap integral) and its comparison with FDTD. Mode index = 1 correlates with the fundamental mode (TE${_0}$ and TM${_0}$). The first 3 mode orders are given for both TE and TM case.}
\label{fig:airmode}
\end{figure}

We now switch the light source from plane-wave input to waveguide light input. In HAMR light delivery, the far-field light is usually delivered by a high-index core material with an 
NFT attached to the waveguide near the ABS. Due to the limited far-field to near-field conversion efficiency, there will be background light leaking from the 
waveguide and reaching the media. This portion of power is not negligible. So although in the theoretical analysis we don't consider the impact of NFT, we think the
general conclusion will still be valuable for understanding the far-field optics of head-media interaction. 

\begin{figure}[htbp]
\centering
\fbox{\includegraphics[width=\linewidth]{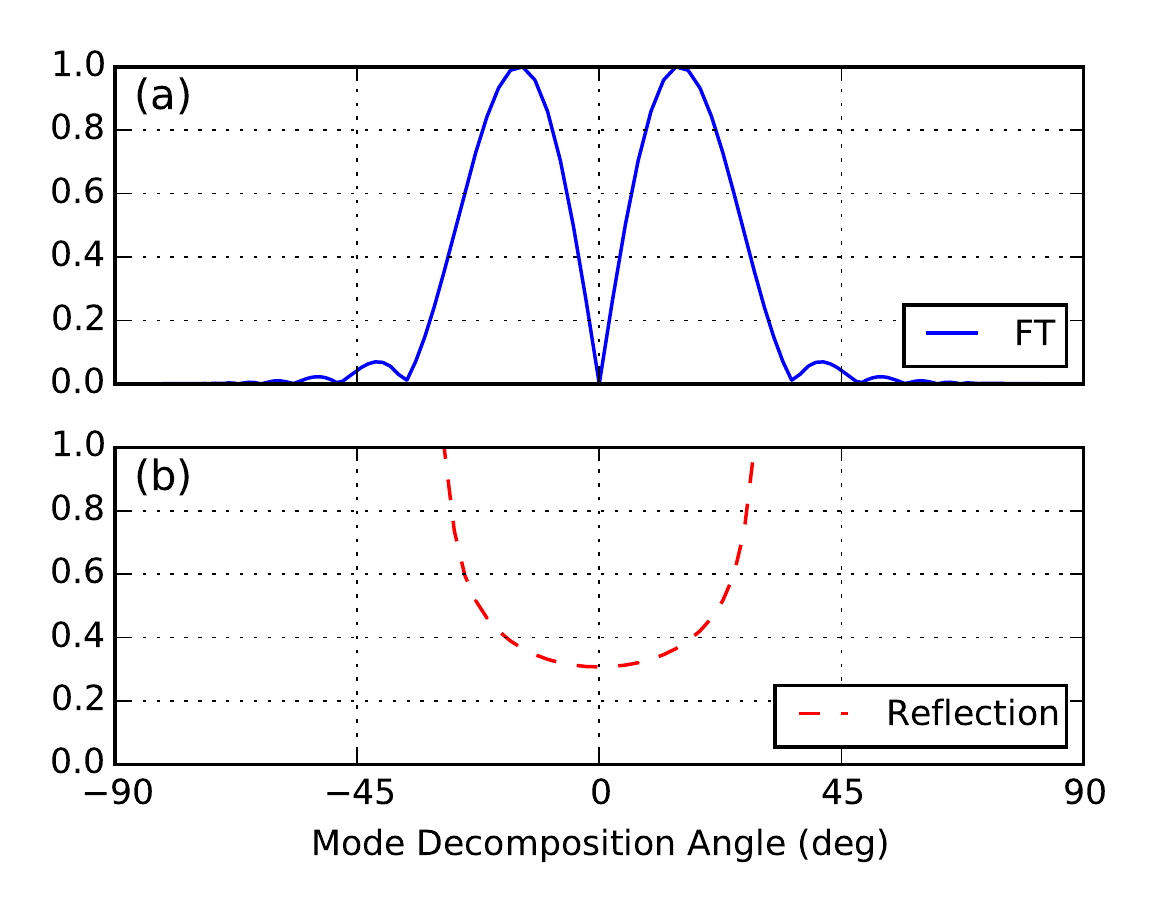}}
\caption{Waveguide mode spectral amplitude and reflection for TE${_1}$ interacting with free-space, with (a) Fourier transform's amplitude and (b) reflection.}
\label{fig:fig5_te10}
\end{figure}

Firstly, we use a multimode waveguide with core index $n_{core}$=2.0, cladding index $n_{clad}$=1.5 and core width $W$ = 800nm as the input and terminate the waveguide in the air ($n$=1.0). The wavelength is chosen as $\lambda$=800nm. Figure \ref{fig:airmode} shows the comparison between PWE and FDTD modeling from Lumerical Solutions \cite{Lumerical}, a commercial-grade simulator based on the finite-difference time-domain method, with decent agreement, validating the PWE with more time-consuming, pure numerical techniques. 

\begin{figure}[htbp]
\centering
\fbox{\includegraphics[width=\linewidth]{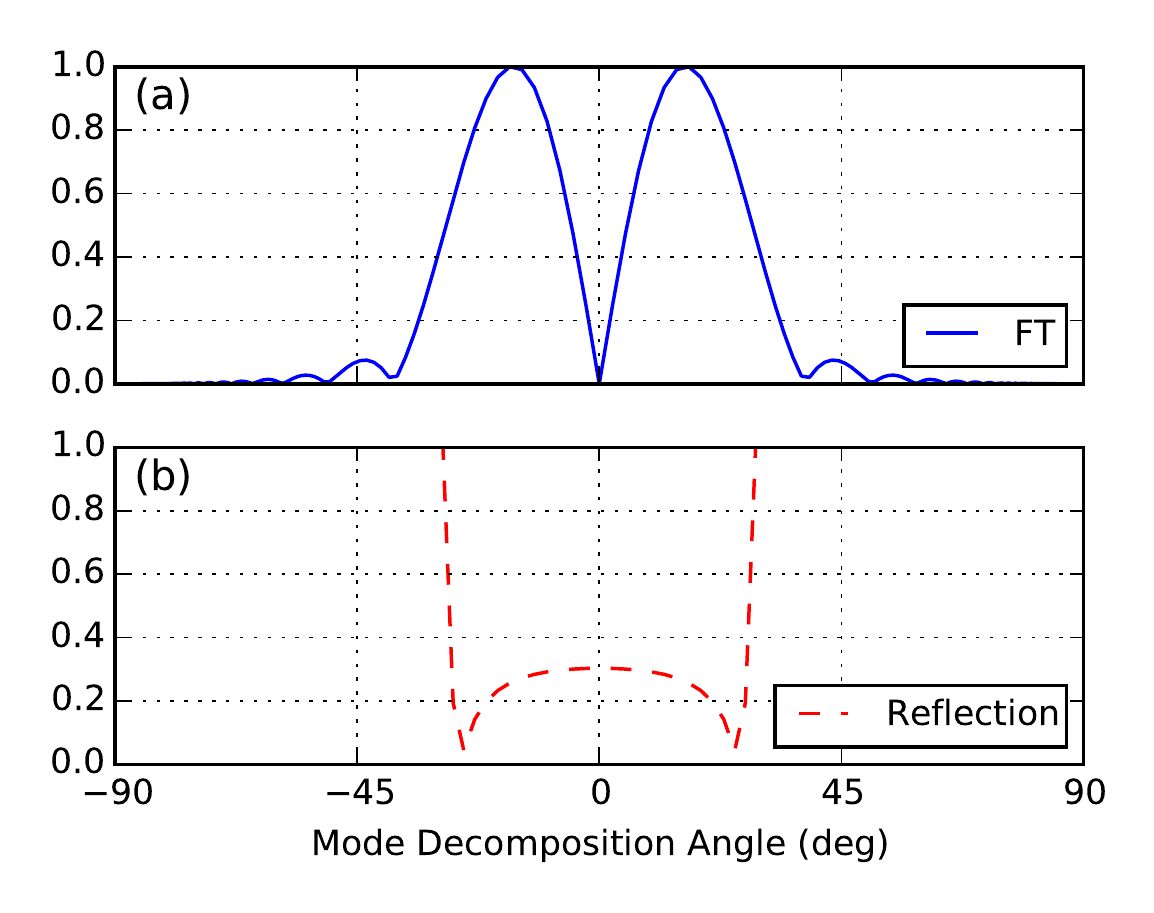}}
\caption{Waveguide mode spectral amplitude and reflection for TM${_1}$ interacting with free-space, with (a) Fourier transform's amplitude and (b) reflection.}
\label{fig:fig6_tm10}
\end{figure}

Choosing the TE${_1}$ and TM${_1}$ mode as a pair of specific case (related to Mode index=2 in Fig. \ref{fig:astcTETM}), we look closer at the spectral map of the waveguide mode and the modal reflection. Figs. \ref{fig:fig5_te10} and \ref{fig:fig6_tm10} show the spectral amplitude (|FT| of the original mode profile) and angular reflection from different polarizations. 

Apparently, on the waveguide-air interface, for TM mode the reflection is generally less. We can see a significant reflection dip at about 25 deg incident angle, which happens to overlap with the Fourier transform peaks of the source field. The reflection dip is a signature of total internal reflection for waveguide-air interface and potentially surface-wave excitation for multilayers with metals. For both situations, the reflection drop will likely contribute to the reduction of back reflection to the waveguide. The surface-wave excitation has been further explored for optical sensing applications, such as in the double metallic cladding waveguide \cite{Li2003}, while waveguide-based Kretschmann configuration has also been studied previously \cite{Shibayama2008Kretschmanntype}. 

\begin{figure}[htbp]
\centering
\fbox{\includegraphics[width=\linewidth]{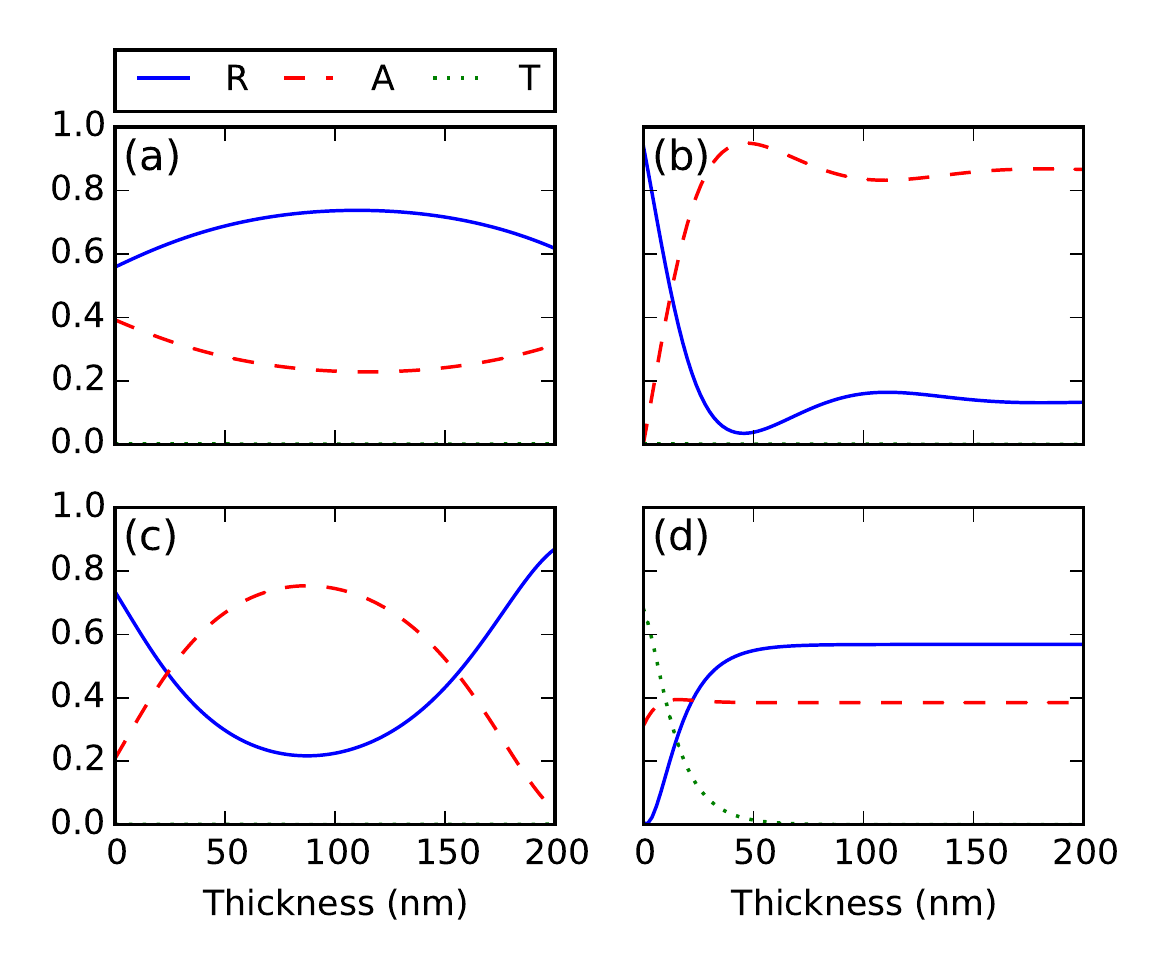}}
\caption{Reflection (labeled 'R', blue solid line), recording layer absorption (labeled 'A', red dashed line) and transmission (labeled 'T', green dotted line) (all normalized to the incident power) for TE${_1}$ vs. different layer thickness, when changing the thickness from 0 to 200nm of (a) overcoat, (b) recording layer, (c) interlayer and (d) heat sink layer.}
\label{fig:swingTE}
\end{figure}

\begin{figure}[htbp]
\centering
\fbox{\includegraphics[width=\linewidth]{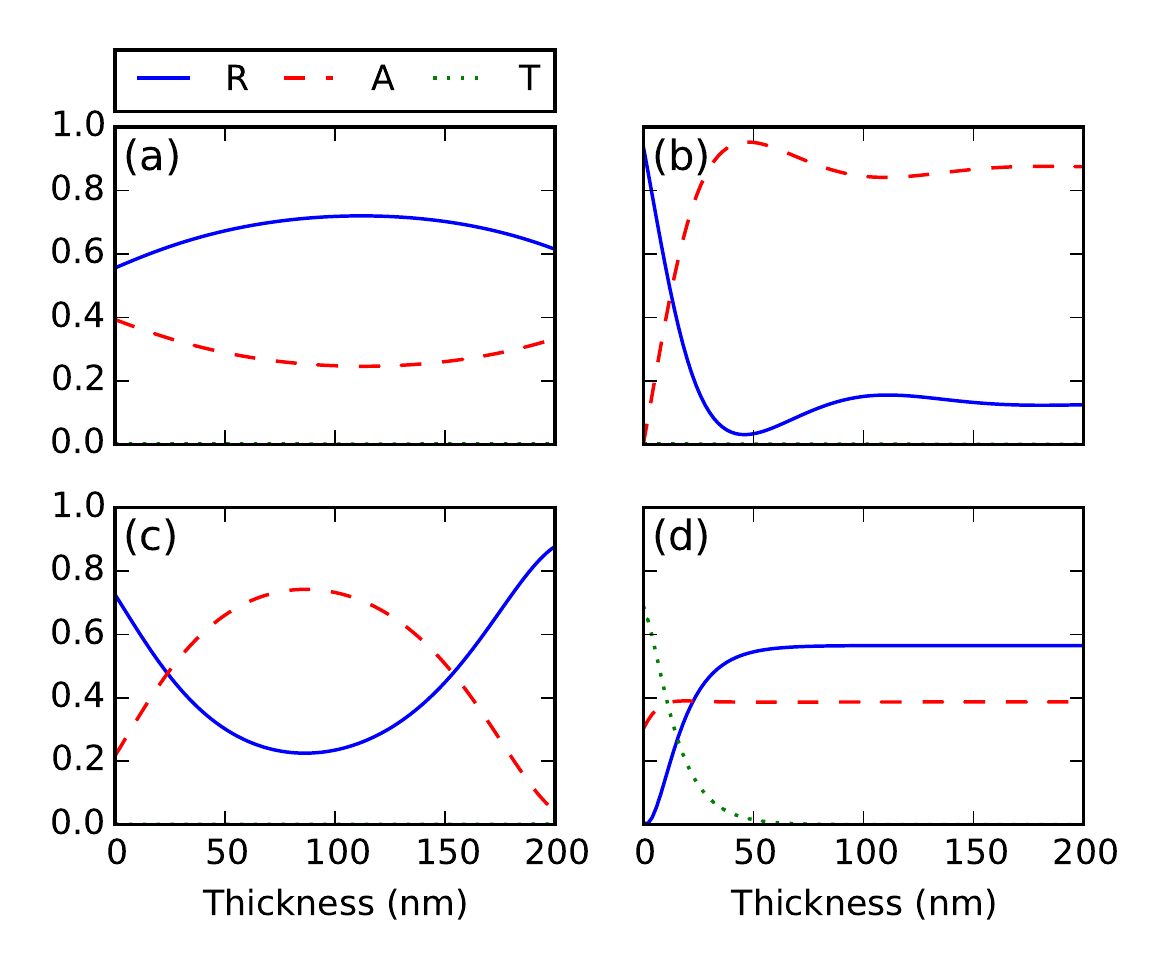}}
\caption{Reflection (labeled 'R', blue solid line), recording layer absorption (labeled 'A', red dashed line) and transmission (labeled 'T', green dotted line) (all normalized to the incident power) for TM${_1}$ vs. different layer thickness, when changing the thickness from 0 to 200nm of (a) overcoat, (b) recording layer, (c) interlayer and (d) heat sink layer. }
\label{fig:swingTM}
\end{figure}

\begin{figure}[htbp]
\centering
\fbox{\includegraphics[width=\linewidth]{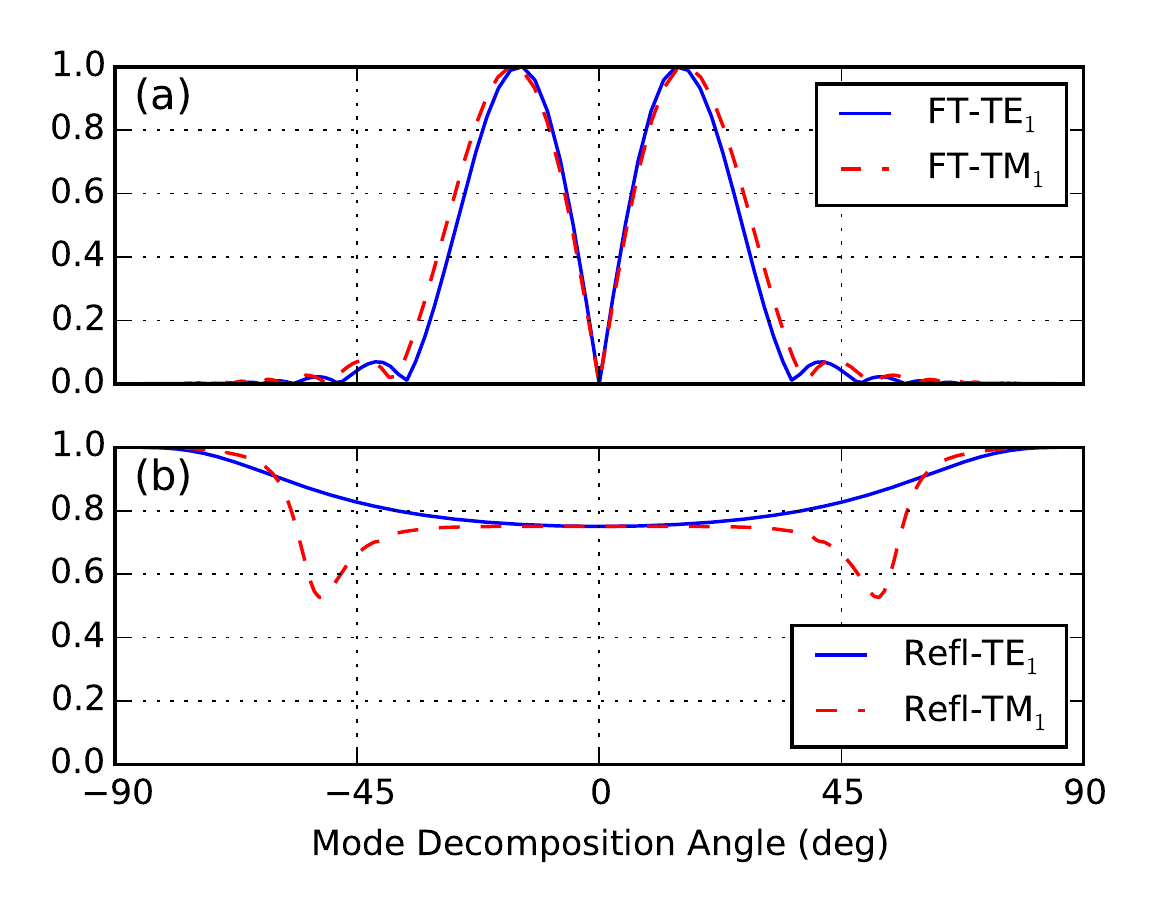}}
\caption{(a) Fourier transform's amplitude and (b) angular reflection comparison for TE${_1}$ (blue solid) and TM${_1}$ (red dashed) waveguide-media interaction.}
\label{fig:astcTETM}
\end{figure}

We then redo the thickness sensitivity study in a similar manner to Figure \ref{fig:swing}, this time with TE${_1}$ and TM${_1}$ modes as the source power. The air gap, recording layer, interlayer and heat sink layer are tuned individually. In Figs. \ref{fig:swingTE} and \ref{fig:swingTM}, we denote the reflection with solid blue curves and absorption with dashed red curves. To observe the curves more closely, we reduce the thickness range from 300nm to 200nm, still including at least a half reflection swing cycle (vs. interlayer thickness). The swing curve for interlayer (as shown in Figs. \ref{fig:swingTE}(c) and \ref{fig:swingTM}(c)) still predicts a reflection valley and a monotonous dropping trend at low interlayer thickness, comparable to the plane-wave case shown in Fig. \ref{fig:swing}. Interestingly, unlike in the waveguide-air case, both TE and TM modes showed similar reflection and absorption for waveguide-media interaction. This effect, seemingly contradiction to the waveguide-air interaction with different polarizations, can be well explained by plotting the angular reflection in a way already seen in Figs. \ref{fig:fig5_te10} and \ref{fig:fig6_tm10}. As shown in Fig. \ref{fig:astcTETM}, the Fourier transform of TE${_1}$/TM${_1}$ modes are fairly close. However, the reflection dip for TM shifts from 25 deg in waveguide-air case, to 50 deg in waveguide-media case. For the angular range containing significant power, which is mostly less than 30 deg, the reflection for TE and TM almost overlaps. It again confirms that HAMR optics should be understood with both head and media in the picture, where changing outlet media could yield significantly different optical behavior with the same waveguide or far-field source.

Finally, we want to point out that Fig. \ref{fig:astcTETM}(b) is the intrinsic angular reflection curve for ASTC stack regardless of the input field. To observe a more profound change between TE and TM mode with this stack, one would need to adjust the waveguide mode profile.

\subsection{HAMR head-media full 3-D Modeling}
The main scope of this paper is to study the far-field light interaction between HAMR head and media. Throughout the theoretical analysis, we didn't consider the role of NFT. Therefore, it is important to check whether the knowledge of far-field light-media interaction is applicable to a complete HAMR system with NFT included, as well as how 2-D learning synchronize with the NFT-media interaction. For HAMR heads, the main values of interest are their far-field reflection (reversely correlated with laser power requirement) and recording performance, where the head reflection and media absorption are optically measurable in 3-D model. We should note again that although the 2-D analysis excludes the NFT, the light interacting with the HAMR media will inevitably include a significant portion of far-field light and there is possibility that the 3-D scenario including both NFT and far-field light be closely related to far-field only situation. In our case, the 3-D model includes an NFT excited by a dielectric light delivery system, as introduced in \cite{Challener2009Heatassisted}.

\begin{figure}[htbp]
\centering
\fbox{\includegraphics[width=\linewidth]{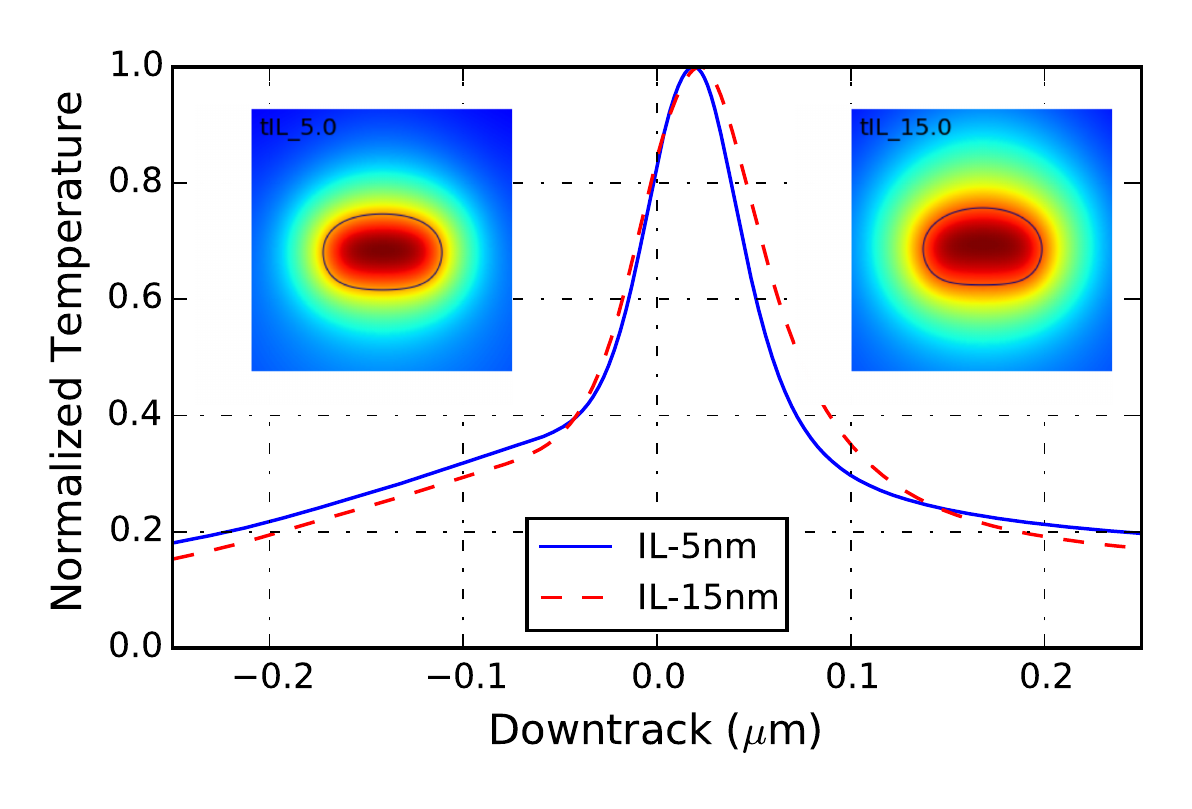}}
\caption{Thermal bubble and downtrack normalized temperature for interlayer thickness at 5nm (blue) and 15nm(red). The inset images show the thermal profile in 2-D for 5nm (left) and 15nm (right) interlayer. }
\label{fig:comsol}
\end{figure}

To start with, we use interlayer thickness as the tuning parameter and present the full 3-D modeling results. In Figure \ref{fig:comsol}, we plot the power scaling factor (normalized to 5nm interlaye case) for a HAMR head. Figure \ref{fig:comsol} also shows the thermal bubble of two cases with 5nm and 15nm interlayer thickness as the insets. The model is built with the RF and heat-transfer modules of COMSOL Multiphysics \cite{Comsol}, using the ohmic heat generation from the RF module as the heat source for the heat-transfer module. Since interlayer is largely a dielectric layer, it is expected that a very thick interlayer will inevitably blow up the writing thermal bubble due to the poorer heat sink efficiency. This has started to show even at very thin interlayer thickness, as the 15nm interlayer showed a locally widened thermal profile compared to 5nm case.

\begin{figure}[htbp]
	\centering
	\fbox{\includegraphics[width=\linewidth]{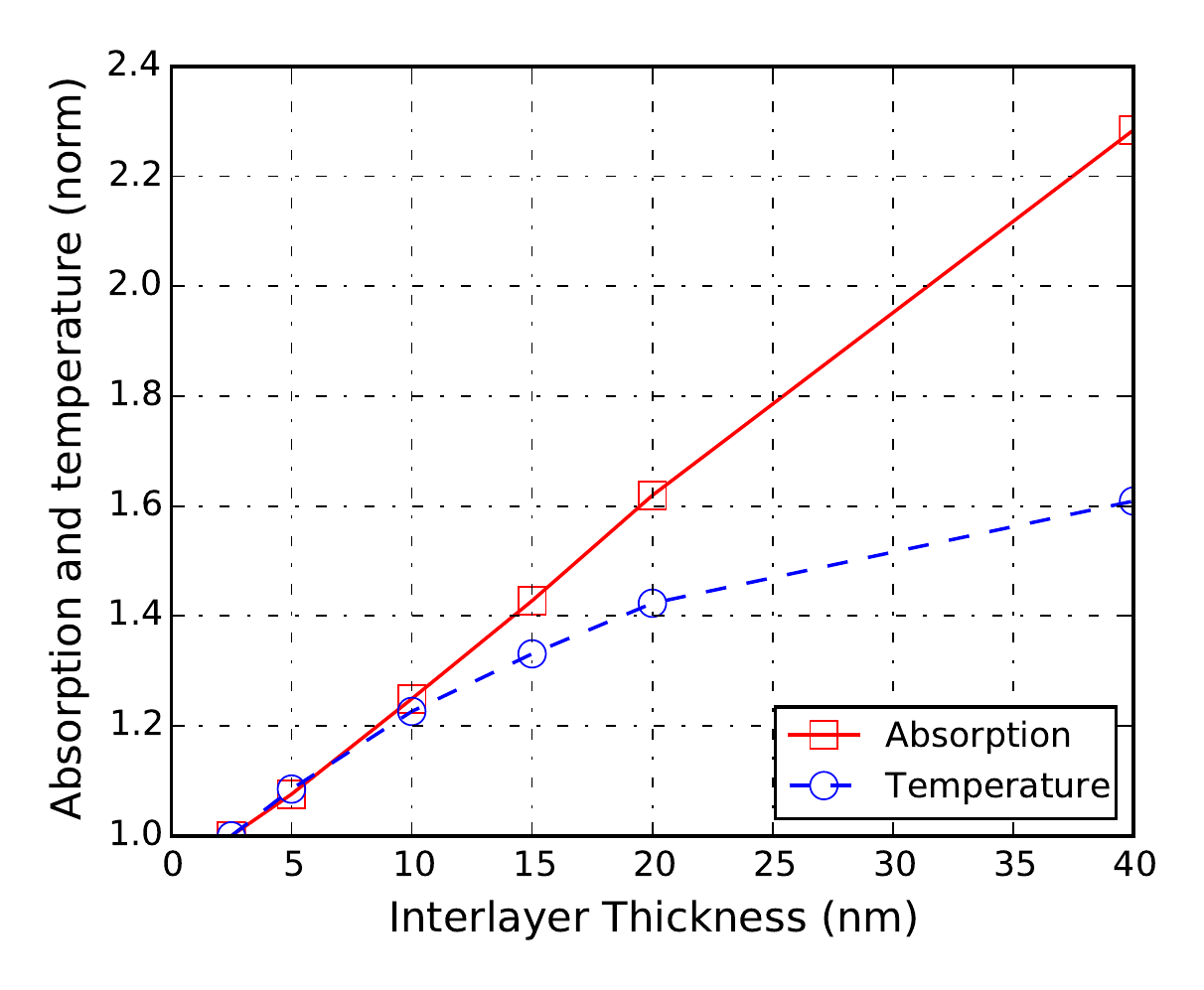}}
	\caption{Interlayer impact on recording layer total absorption and 
		peak temperature.}
	\label{fig:abs3d}
\end{figure}

On the other hand, Figure \ref{fig:abs3d} shows how absorption (of magnetic layer) and modeled magnetic layer peak temperature change with interlayer thickness. We see that although in 3D model there are both near- and far-field optical interaction, the general trend predicted previously for reflection and absorption still qualitatively applies. This suggests that either the near-field light induces similar media sensitivity for magnetic layer absorption, or the far-field proportion dominates the large-scale heat generation and temperature rise. While the near-field light-media interaction is out of the scope of this paper, we think it is possible that both factors mentioned above play a role in this observation. For the temperature curve, it is worthy of mentioning that the increased interlayer will decrease the overall heat sink efficiency of the stack and as a result preserve more heat in recording layer better. Since the temperature curve is bending toward a flatter slope, we think quantitatively the thermal isolation role played by thicker interlayer is less substantial compared to its optical or impedance matching role.

The models given here will be the reference point of the experiments summarized in the next section.

\section{Experimental results}
\label{sec:experimental}

In this section, we focus on the experimental verification of previous modeling data. Specifically, from the large sensitivity slope in interlayer thickness, we expect to see significant laser diode current (LDC) change with as small as 10nm interlayer thickness change. Based on this, we prepared four flyable HAMR media discs designed and fabricated with different interlayer thickness (from 2.5nm to 15nm) and performed recording tests on two of them (Media A5 with thickness $d$=5nm and media A15 with $d$=15nm). We want to emphasize that the actual physical properties of the fabricated media stacks understandably deviates from the generic media in Table \ref{tab:astc} partly due to process variations, but the main component remains physically consistent compared to the generic stack, where we obtained most of the thickness sensitivity learning. 

\begin{figure}[htbp]
	\centering
	\fbox{\includegraphics[width=\linewidth]{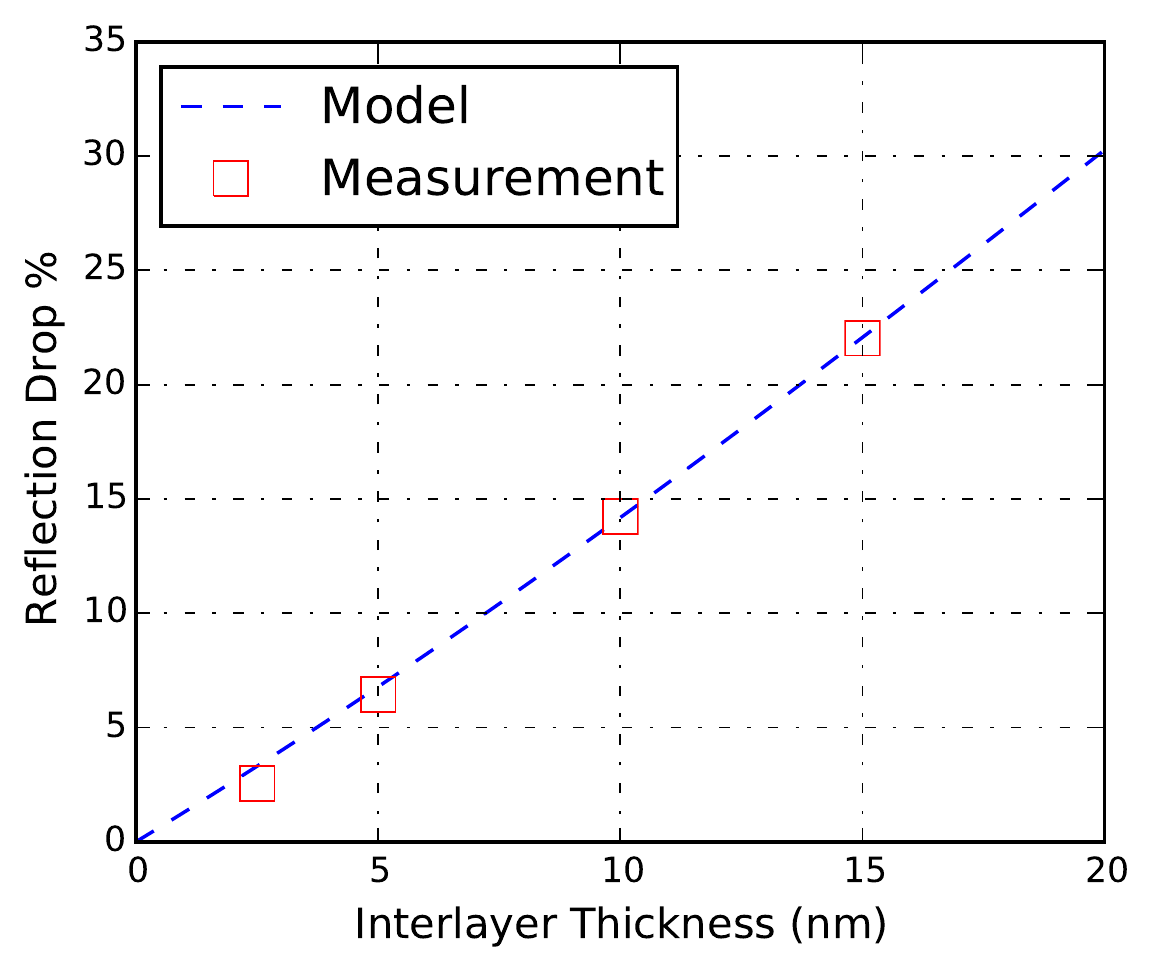}}
	\caption{Far-field media reflection drop vs. interlayer thickness change.}
	\label{fig:ladder}
\end{figure}

First we obtained the far-field reflection of the interlayer ladder. Figure \ref{fig:ladder} shows the modeled and measured media reflection with decent agreement (normalized to the largest modeled reflection, when modeled interlayer thickness $d$ approaches 0). Note that the y-axis of the plot is the relative reflection-drop, compared to zero thickness case (modeled) which has the maximum reflection. From Figure \ref{fig:swing} we have noticed that for thin interlayer ($<$50nm), the reflection changes with interlayer thickness monotonously. If the overall heating process combining near- and far-field light followed similar sensitivity trend to the far-field-only case discussed above, we expect to see $>$30\% laser power drop with only 10nm interlayer thickness change.

For recording performance evaluation, we use a number of HAMR heads to write on both Media A5 and A15, so as to obtain the LDC and recording bit-error-rate (BER). 

\begin{figure}[htbp]
	\centering
	\fbox{\includegraphics[width=\linewidth]{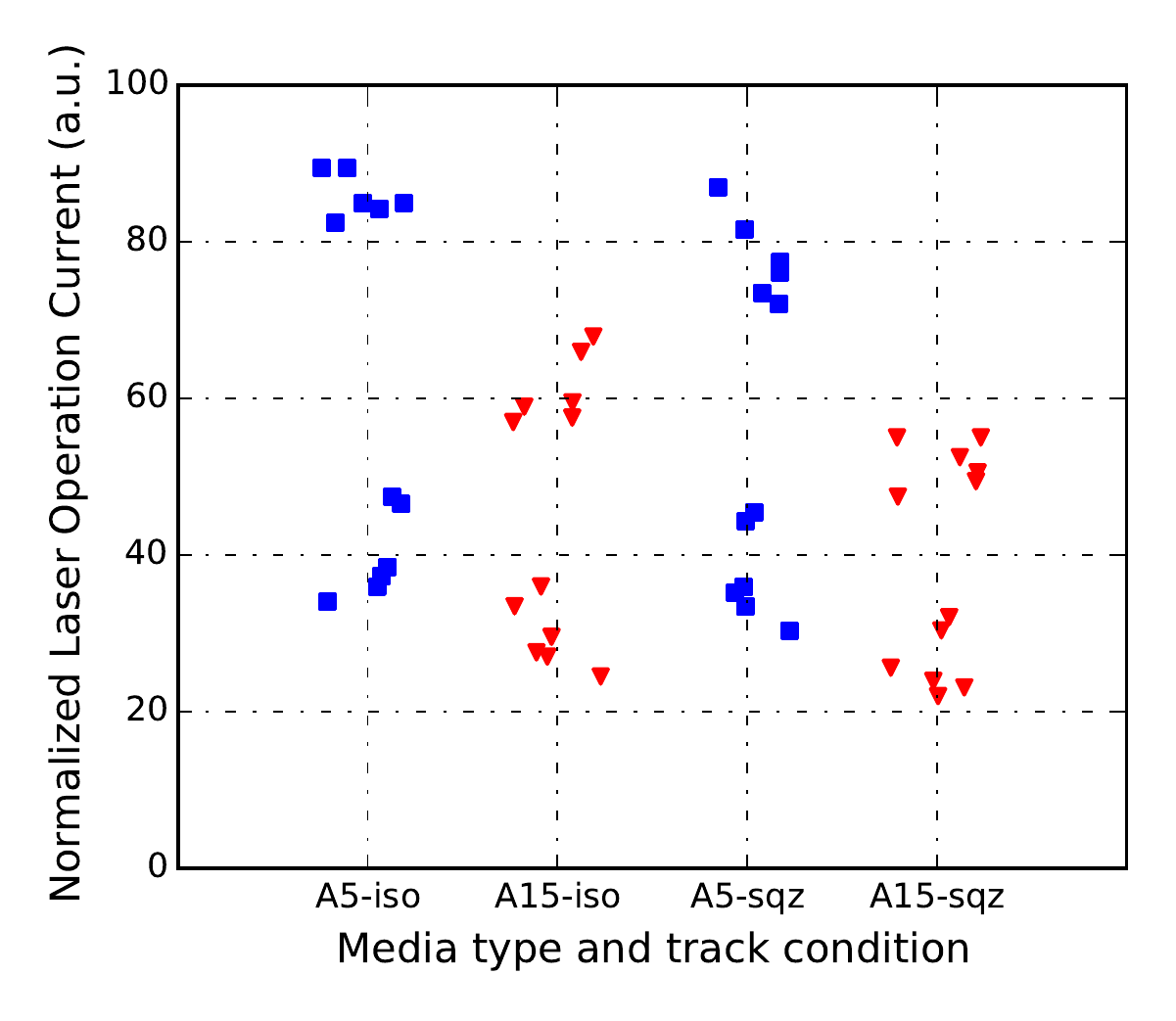}}
	\caption{Laser power requirement (normalized) with A5/A15 under different writing conditiosn, iso=single track, sqz=multiple track.}
	\label{fig:lp}
\end{figure}

Figure \ref{fig:lp} showed the operational currents (subtracting the threshold current from LDC before normalization to a bench-mark value) in 4 writing conditions. While A5 and A15 denote the media difference, "iso" and "sqz" describes the single- or multiple-track recording conditions, on which the laser power is determined with different optimization approach. Apparently, between "iso" and "sqz" the laser power difference is generally minor for both media. On the other hand, the laser power requirement to write A15 is about 25\%-30\% lower compared to A5, indicating a higher heating efficiency on A15 with similar heads on. The agreement between the experiments and the far-field-only optical behaviors are notable. We want to point out again that the drop of LDC is an result combining the optical and thermal effects from thicker interlayer thickness, as already discussed around Fig. \ref{fig:abs3d}.

\begin{figure}[htbp]
\centering
\fbox{\includegraphics[width=\linewidth]{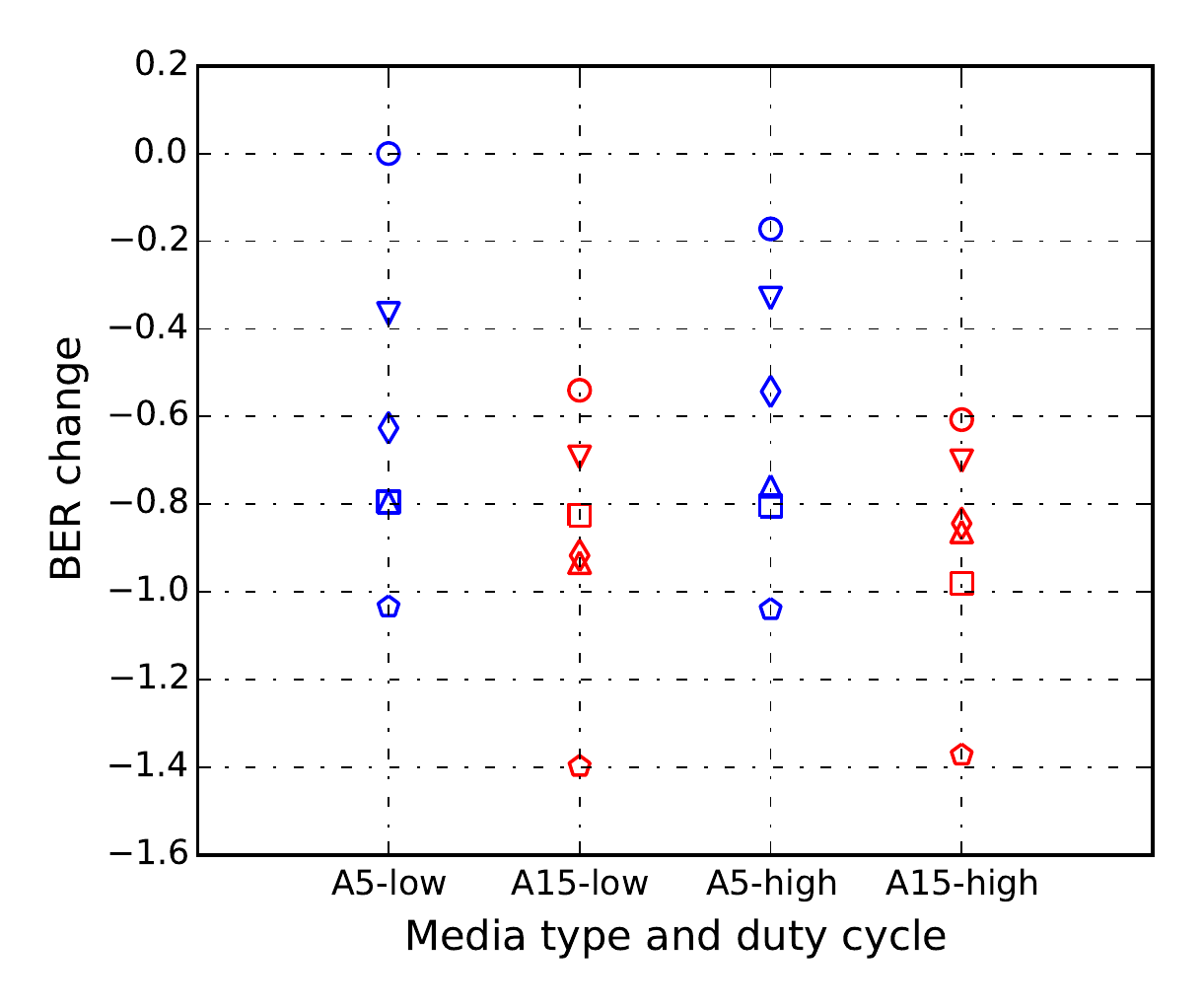}}
\caption{BER change (relative to the worst measured BER) with A5/A15 under different writing conditions, low=low duty cycle, high=high duty cycle. The same marker indicates the same head used for recording.}
\label{fig:ber}
\end{figure}

Figure \ref{fig:ber} summarized the BER change of recording tests on A5 and A15 at two different laser on-off duty cycles, relative to the worse BER seen in the experiments. There are 6 heads in total tested on both media, and we find the laser duty cycles do not change the BER significantly. Similar to the LDC comparison above, it is the switch from A5 to A15 that has the most profound impact on BER. While A15 did show a better recording performance for the single-track writing, it is expected that a worse heatsinking efficiency bounded with thicker interlayer thickness will ultimately produce blown-up thermal bubbles that makes high areal density recording impossible.

The observation shown in this section confirms that the far-field light-media interaction is highly relevant to HAMR recording, at least for the light delivery system studied in this paper. The similar sensitivity shown between waveguide-only in 2D as shown in section \ref{sec:modeling} and the full-3D case in this section suggests that under certain situation, the near-field heating process does not behave significantly different from its far-field counterpart. On the other hand, the HAMR heating is known to have both near- and far-field contribution. A far-field-like heating source, such as the waveguide radiation directly emanating from a channel or slab waveguide to HAMR media, provides background lighting and increases the thermal background's intensity. From a contrast point of view, this background generally lowered the relative intensity and sharpness of the near-field heating spot. It is expected that towards the direction of more refined HAMR optical system, one can reduce the background light by either trimming or blocking the waveguide light and suppress the back-reflection by engineer the mode profile to match with the media.

\section{Conclusion}

We have used a plane wave expansion method to study the waveguide-media optical interaction in HAMR technology, and noted the media sensitivity regarding the individual layer's optical thickness. The sensitivity predicted by model is verified by writing tests, showing the importance of matching the optical system with media design.

\section*{Acknowledgments}

R.Y. acknowledges the discussion with Dr. Thach Nguyen of RMIT, Australia on technical details of implementing the waveguide plane wave expansion.

\bigskip



\end{document}